\documentclass[12pt]{article}
\usepackage{txfonts}
\usepackage[letterpaper,portrait,margin=1in]{geometry}
\usepackage{graphicx}
\usepackage{wrapfig}
\usepackage[utf8]{inputenc}
\usepackage{enumitem,amssymb}
\usepackage{ragged2e}
\usepackage{pifont}
\usepackage[round,sort]{natbib}
\usepackage{xspace}
\usepackage[hidelinks]{hyperref}
\usepackage{aas_macros}
\usepackage{paralist}
\usepackage{microtype}
\usepackage{multicol}
\usepackage{booktabs}
\usepackage{tabu}

\newlist{thematic}{itemize}{8}
\setlist[thematic]{label=$\square$}

\newcommand{\nustar}{\textit{NuSTAR}\xspace}

\newcommand{\strobex}{\textit{STROBE-X}\xspace}

\makeatletter
\renewcommand\section{\@startsection {section}{1}{0mm}%
                     {-0.2\baselineskip}%
                     {0.1\baselineskip}%
                     {\normalfont\large\bfseries}}%
\renewcommand{\subsection}{\@startsection {subsection}{2}{0mm}%
                          {-0.2\baselineskip}%
                          {0.1\baselineskip}%
                          {\normalfont\bfseries}}
\renewcommand{\subsubsection}{\@startsection {subsubsection}{3}{0mm}%
                          {-0.2\baselineskip}%
                          {0.1\baselineskip}%
                          {\normalfont\bfseries}}%
\makeatother

\begin{document}

\noindent {\huge Astro2020 Science White Paper \\
The Physics of Accretion Onto Highly \\
Magnetized Neutron Stars \par}

\vspace*{\baselineskip}

\noindent \textbf{Thematic Area:} Formation and Evolution of Compact Objects 

\vspace*{\baselineskip}
  
\noindent \textbf{Principal Author:}
\noindent {{\bf Michael T. Wolff}, Space Science Division, Naval Research Laboratory,\\ 
Washington DC 20375,  USA, 
Email: Michael.Wolff@nrl.navy.mil, 
Phone: (202) 404-7597}

\vspace*{\baselineskip}
 
\noindent \textbf{Co-authors:} \\
{\bf Peter A. Becker}, George Mason University, Fairfax, VA 22030,  USA\\
{\bf Joel Coley}, Howard University, Washington, DC, 20059, USA\\
{\bf Felix F\"urst}, European Space Astronomy Centre, Villanueva de la Ca\~nada, Madrid, Spain\\
{\bf Sebastien Guillot}, Institut de Recherche en Astrophysique et Planétologie, Toulouse, France\\
{\bf Alice K. Harding}, NASA Goddard Space Flight Center, Greenbelt, MD 20771, USA\\
{\bf Paul B. Hemphill}, MIT Kavli Institute, Cambridge, MA 02139, USA\\
{\bf Gaurava K. Jaisawal}, National Space Institute, DK-2800 Lyngby, Denmark\\
{\bf Peter Kretschmar}, European Space Astronomy Centre, Villanueva de la Ca\~nada, Madrid, Spain\\
{\bf Matthias Bissinger n{\'e} K\"uhnel}, ECAP, FAU Erlangen-N\"urnberg, 96049 Bamberg, Germany\\
{\bf Christian Malacaria}, National Space Science and Technology Center, Huntsville, AL 35805, USA\\
{\bf Katja Pottschmidt}, University of Maryland Baltimore County, Baltimore, MD 21250, USA\\
{\bf Richard E. Rothschild}, CASS, University of California San Diego, La Jolla, CA 92093, USA\\
{\bf R\"udiger Staubert}, IAA, University of T\"ubingen, 72076 T\"ubingen, Germany\\
{\bf John A. Tomsick}, University of California at Berkeley, Berkeley, CA 94720, USA\\
{\bf Brent West}, Commander, Naval Surface Force, U.S. Pacific Fleet, San Diego, CA 92155, USA\\
{\bf J\"orn Wilms}, ECAP, FAU Erlangen-N\"urnberg, 96049 Bamberg, Germany\\
{\bf Colleen Wilson-Hodge}, NASA Marshall Space Flight Center, Huntsville, AL 35812, USA\\
{\bf Kent S. Wood}, Praxis, Inc., Alexandria, VA 22303, USA

\vspace*{\baselineskip}

\noindent {\textbf{Abstract:} Studying the physical processes occurring in the region 
just above the magnetic poles of strongly magnetized, 
accreting binary neutron stars is essential to our understanding 
of stellar and binary system evolution. 
Perhaps more importantly, it provides us with a natural laboratory 
for studying the physics of high temperature and high density plasmas 
exposed to extreme radiation, gravitational, and magnetic fields. 
Observations over the past decade have shed new light on the manner 
in which plasma falling at velocities near the speed of light onto 
a neutron star surface is halted. 
Recent advances in modeling these processes have resulted in direct 
measurement of the magnetic fields and plasma properties. 
On the other hand, numerous physical processes have been identified that
challenge our current picture of how the accretion process onto neutron stars works. 
Observation and theory are our essential tools in this regime because the 
extreme conditions cannot be duplicated on Earth. 
This white paper gives an overview of the current theory, the outstanding 
theoretical and observational challenges, and the importance of 
addressing them in contemporary astrophysics research.}
\pagebreak

\section{Motivation} \label{sec:motivation}
Neutron stars (NSs) occupy a unique position in the Universe. 
They are the only laboratories for the study of cold, ultradense 
matter, likely exceeding nuclear densities at their 
cores \citep[see, e.g.,][for a review]{lattimer_2012}. 
They can harbor extremely strong magnetic fields, up to $10^{15}$\,G, 
representing the only place in the Universe where we can study such fields. 
And, as GW170817 \citep{abbott_multimessenger_2017} has shown us, NS mergers 
are the primary source of heavy elements in the Universe. 
Hence, studying NS physics will remain a major research priority 
in the coming decade.

Neutron star physics can be ideally studied in accreting binary 
systems, where the NS accretes matter from its less-evolved companion. 
Systems with massive donors 
\citep[``high-mass X-ray binaries'' or HMXBs, with OB-star donors;][]{meszaros_high-energy_1992} 
are of special interest as possible NS-NS merger progenitors. 
The accretion by the NS is a key element to understanding the 
evolution of these systems. Furthermore, binary systems containing 
a NS and a Be-type companion are also of special interest, since 
these systems can display highly variable accretion rates onto the NS.

\begin{wrapfigure}{R}{0.40\textwidth}
\includegraphics[width=0.38\textwidth]{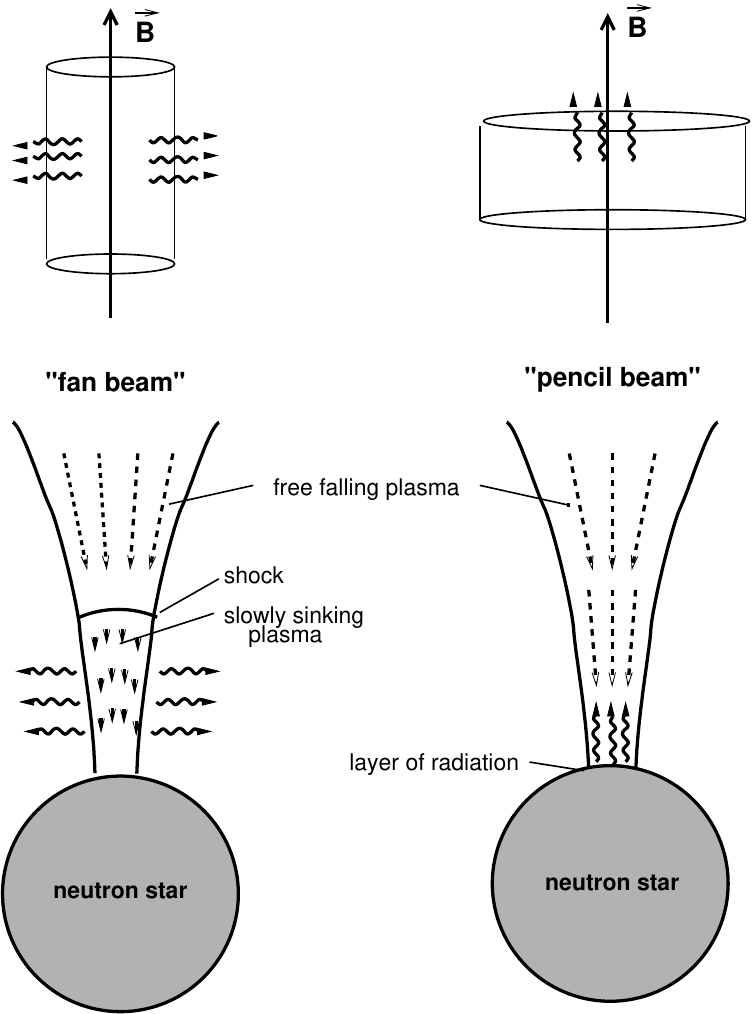}
\caption{ 
Left: ``Fan-beam''/cylinder geometry. 
Right: ``pencil beam''/slab geometry. From \citet{schonherr_model_2007}.}
\label{fig:AccretionGeom}
\end{wrapfigure}

An accreting NS gravitationally captures matter from its companion, 
either via a stellar wind, Roche lobe overflow, or from the disk 
of a Be-type companion. 
This often forms an accretion disk around the NS, similar to what 
is seen in black hole (BH) binaries \citep[e.g.,][]{shakura_black_1973,ghosh_disk_1978}.
However, the NS's extremely strong magnetic field disrupts the disk at the 
magnetospheric radius, $R_{m}{\sim}10^8$\,cm, where the magnetic pressure 
begins to dominate over the ram pressure of the infalling material. 
The disk inner edge is thus significantly farther out than in a BH, 
and the fluid-like mixture of ionized gas, coupled with radiation, 
is channelled along the field lines into the top of the NS accretion 
column over the NS polar cap region, reaching free-fall velocities 
of ${\sim}0.5c$ with ion and electron temperatures up 
to $10^7$-$10^8$\,K \citep[see, e.g.,][]{basko_limiting_1976,west_dynamical_2017}.

The disk properties, and the disk truncation distance, are 
mass transfer rate and NS magnetic field strength dependent. 
While there are a number of methods for estimating NS magnetic 
field strengths, many are indirect and rely on poorly-understood 
phenomena. 
However, accreting NSs provide one of the few \textit{direct} 
measurements of the field strength: \textit{cyclotron 
resonant scattering features} (CRSFs, or ``cyclotron lines''). 
Additionally, as these features are produced by photons 
scattering on electrons in ${\sim}10^{12}$\,G fields, they act as 
probes of the environment close to the NS surface. 
CRSFs make accreting NSs highly attractive targets for detailed 
studies of accretion. 
However, in order for this information to be useful, we must 
have a thorough knowledge of the physical processes involved, 
as well as a wide-ranging sample of data with which to test our understanding.

\section{Physics of Emission from the Magnetic Poles of Neutron Stars} \label{sec:physics}
As ionized matter falls towards the surface of the star, soft X-ray photons generated 
in the flow 
are Compton upscattered in energy, cooling and 
braking the plasma while producing the observed hard X-rays 
\citep[][and references therein]{becker_thermal_2007,west_dynamical_2017}. 
The observed spectrum typically resembles a power-law with energy at low energies, 
transitioning to an exponentially falling spectrum above 10$-$20\,keV. 
Analytical and numerical treatments of this ``Comptonization'' process and the 
resulting X-ray spectra have been presented, 
\citep[e.g.,][]{becker_thermal_2007,farinelli_numerical_2012,postnov_dependence_2015,west_spectra_2017}. 
The response of the flow to this cooling and braking, and to changes in the 
accretion rate, $\dot{M}$, was studied by \citet{becker_spectral_2012}, who 
found that the behavior of the column is determined by two main factors: is  
the infalling material moving supersonically? And, does radiation pressure 
play a significant role in halting the flow? 
Figure~\ref{fig:AccretionGeom}, from \citet{schonherr_model_2007}, displays two 
extremes of behavior: the left side shows the high 
$\dot{M}$ case ($\dot{M}{\gtrsim}10^{17}$\,g\,s$^{-1}$), with 
supersonic flow, significant radiation pressure, and ``fan-beam'' emission 
from the sides of the column, while the right side shows the low $\dot{M}$ 
case ($\dot{M}{\sim}10^{14}$--$10^{15}\,\mathrm{g}\,\mathrm{s}^{-1}$), where the 
infalling matter impacts directly on the NS surface and is halted by Coulomb 
collisions, producing a so-called ``pencil-beam'', where the emission 
is directed mostly upwards, along the field lines.

\subsection{Cyclotron Resonance Scattering Features – Measuring the Magnetic Field} \label{sec:crsf}
\begin{figure}
    \hspace*{2em}
    \includegraphics[height=11\baselineskip]{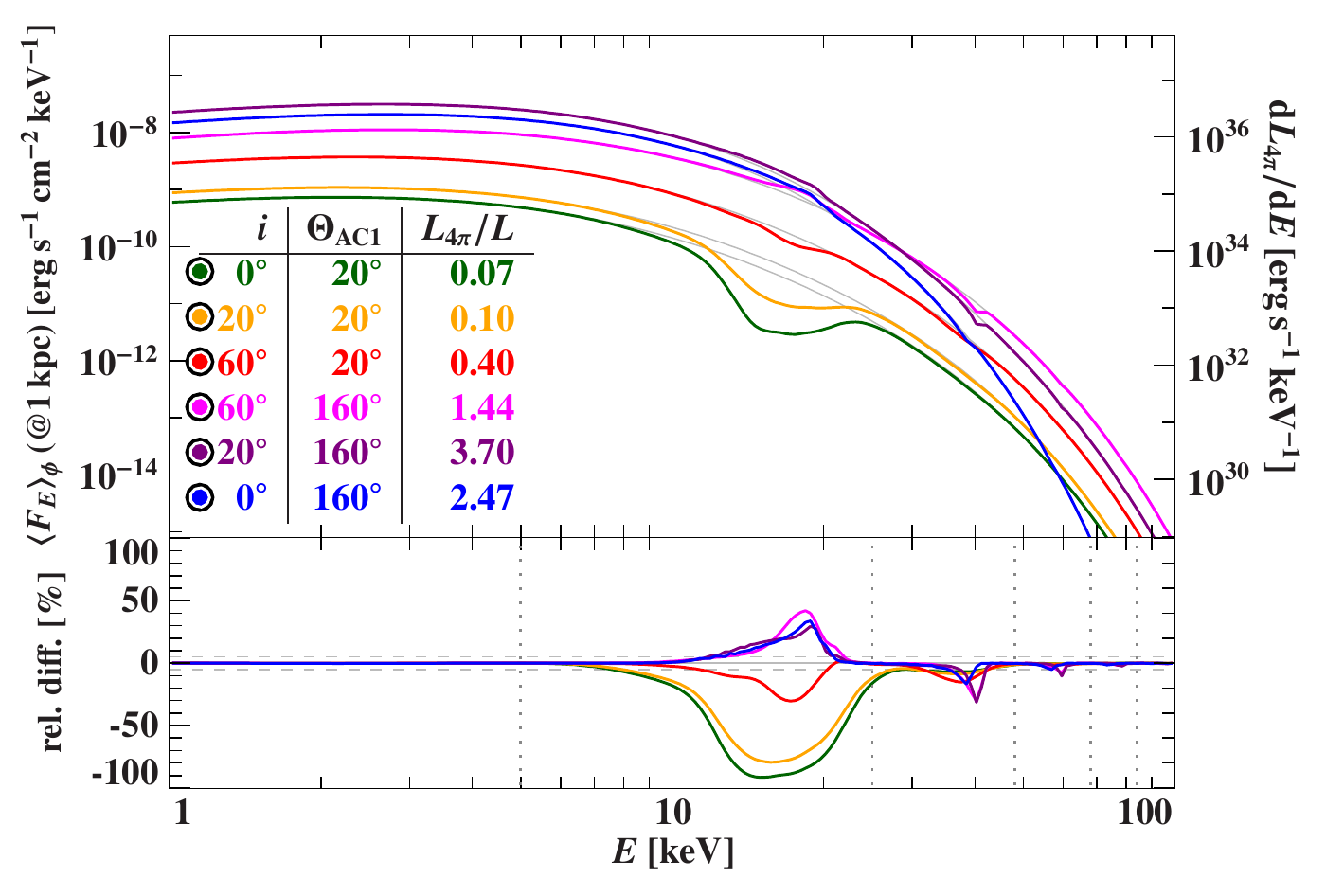}
    \hfill
    \includegraphics[height=11\baselineskip]{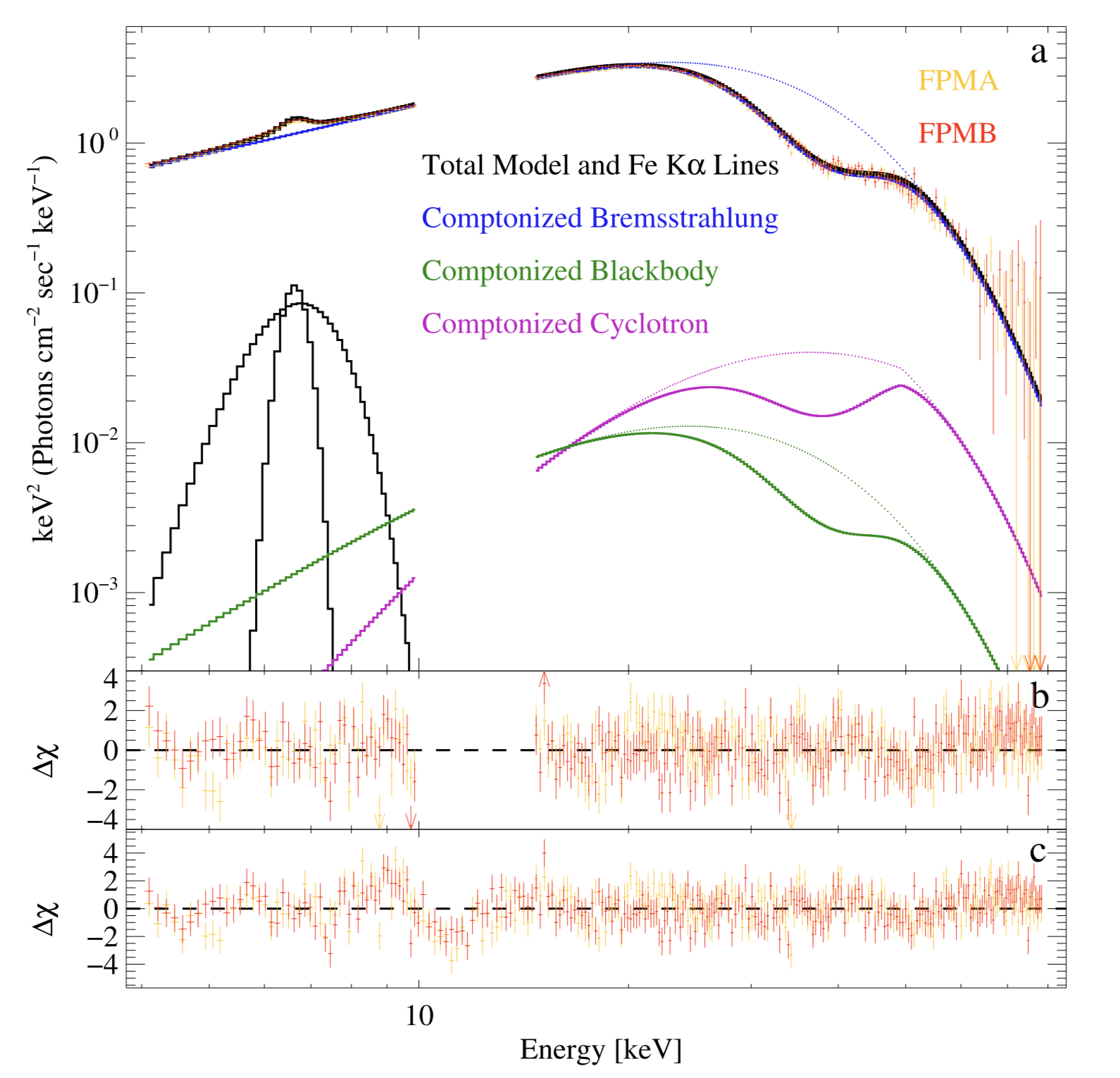}\hspace*{2em}
    \vspace{-2ex}
    \caption{\textit{Left:} Simulated observed spectra from a NS accretion column at different viewing angles, from S. Falkner et al. (submitted). \textit{Right:} From \citet{wolff_nustar_2016-1}, the observed \nustar spectrum of Her~X-1 fitted with the radiation-dominated radiative shock model of \citet{becker_thermal_2007}.}
    \label{fig:sim_spectra}
\vspace{-2ex} 
\end{figure}

In addition to the cut-off power-law X-ray spectrum described above, 
approximately three dozen of the $\sim$350 known accreting pulsars 
also display broad absorption-like features in the 10--100\,keV 
range \citep{staubert_cyclotron_2019}. 
These \textit{cyclotron resonance scattering features} are due 
to resonant scattering of photons on electrons in the strong 
magnetic field. 
The perpendicular motion of electrons relative to magnetic fields 
is quantized into Landau levels, the approximate spacing of 
which is given by 
$\Delta E \approx 11.6 \,{\rm (keV)} \, B_{12}$, where $B_{12}$ is the 
field strength in $10^{12}$\,G. 
This results in a highly energy- and direction-dependent 
scattering cross-section, with resonances at (approximately) 
integer multiples of the energy spacing $\Delta E$.  
NSs with field strengths of $10^{12}$--$10^{13}$\,G thus display 
these features in the 10--100\,keV band, accessible to hard 
X-ray spectrometers (see Fig.~\ref{fig:sim_spectra}, right). 
The observed features are broad ($\Delta E/E{\sim}0.1$), likely 
due to a combination of thermal motion and 
averaging over different angles to the magnetic field. 
The dependence on angle also means these features are 
intrinsically dependent on the pulse phase of the observed 
pulsar, reinforcing the need for missions with the time 
resolution and effective area to carry out pulse phase-resolved 
analyses. 

Building on earlier work by \citet{wang_cyclotron_1988} and \citet{araya_cylotron_1996}, 
we are now able to calculate self-consistent line profiles for cyclotron lines 
\citep{schonherr_model_2007,schonherr_formation_2014,
schwarm_cyclotron_2017a,schwarm_cyclotron_2017b}, 
and by combining these results with new models for the X-ray continuum 
(see Section~\ref{sec:continuum} below) and gravitational light bending, we 
can produce realistic X-ray spectra and pulse profiles (Falkner et al., 
submitted, see Fig.~\ref{fig:sim_spectra}).
Comparing these models to observations is difficult, due to the computational 
expense and intrinsically large number of free parameters, but not impossible, and 
advances in computing power and analysis techniques (e.g., Markov Chain Monte 
Carlo analyses or machine learning approaches) will make this a 
significantly more feasible goal in the coming decade. 
This goal must be pursued if we want to further our
understanding of the extreme physical processes occurring in the NS 
environment, which cannot be duplicated on Earth.

\subsection{Detailed Modeling of the Continuum from First Principles} \label{sec:continuum}
The extreme physics of the flow structure in the column makes 
modeling the continuum generated by accretion onto the magnetic 
poles of NSs very difficult.
The radiation-dominated flow model \citep{davidson_accretion_1973,arons_radiation_1987} 
has shown considerable promise in accounting 
for the radiation properties of luminous accreting X-ray pulsars such as Her~X-1. 
We already know that unlike the Coulomb collisional stopping 
and collisionless shock models \citep[e.g.,][]{langer_low-luminosity_1982}, 
radiation-dominated flow models can account for the general 
shape of the X-ray spectra at X-ray luminosities near and above 
the ``critical luminosity'' \citep[][see Fig. 2]{wolff_nustar_2016-1}. 
The theoretical spectra display exponential Compton cut-offs in 
the high energy X-ray range (10--40\,keV) just like those observed, and power-law 
continua that can be understood as Compton upscattering of 
the high energy photon distributions in the plasma. 
This theoretical spectrum shows agreement with the observed 
spectrum over nearly 2 orders of magnitude in energy.

A critical assumption in the emerging theoretical development, 
and one that is almost certainly incorrect, is that one type of 
flow model, namely a radiation-dominated shock where electron 
scattering is completely efficient in stopping the plasma 
flow near the NS surface, is applicable in all sources, from 
the low luminosity steep-spectrum sources such as X Persei, 
to the high luminosity flat spectrum sources such as LMC~X-4
and Cen~X-3. 
But how do accretion flows onto NSs transition from 
radiation-dominated to gas-dominated? This is not known. 
Perhaps as the luminosity decreases, gas-mediated, thermal 
``sub-shocks'' develop in the radiation-dominated flows in 
a manner suggested by \citep{becker_exact_2001} 
in their study of cosmic-ray acceleration in 
supernova shock waves. 
Another critical question for our understanding of 
the accretion flow structures is whether the emergent 
radiation comes out in a fan beam, in a pencil beam, or both?

Another aspect of the problem is the energy budget. 
This involves the details of the production of the cyclotron, 
bremsstrahlung, and blackbody ``seed photons,'' and how these 
photons interact with the gas to heat or cool the plasma, 
extract energy from the accretion column, and form the 
broad continuum spectrum. In principle, the dynamics and the 
spectral formation must be treated self-consistently, which 
is an extremely difficult ``grand challenge'' level simulation. 
Initial steps in this direction have been accomplished 
by \citet{west_dynamical_2017,west_spectra_2017}, but 
these calculations were limited to one spatial direction (radial), 
and are not time-dependent. 
In a fully self-consistent model, the radiation regulates the 
structure of the flows as it 
merges with the NS atmosphere. 

\section{Recent Observational Advances and Goals for the Coming Decade} \label{sec:observe}

\subsection{The CRSF Energy-Luminosity Relationship} \label{sec:crsf_lum}

\begin{figure}
\centering
\includegraphics[height=9\baselineskip]{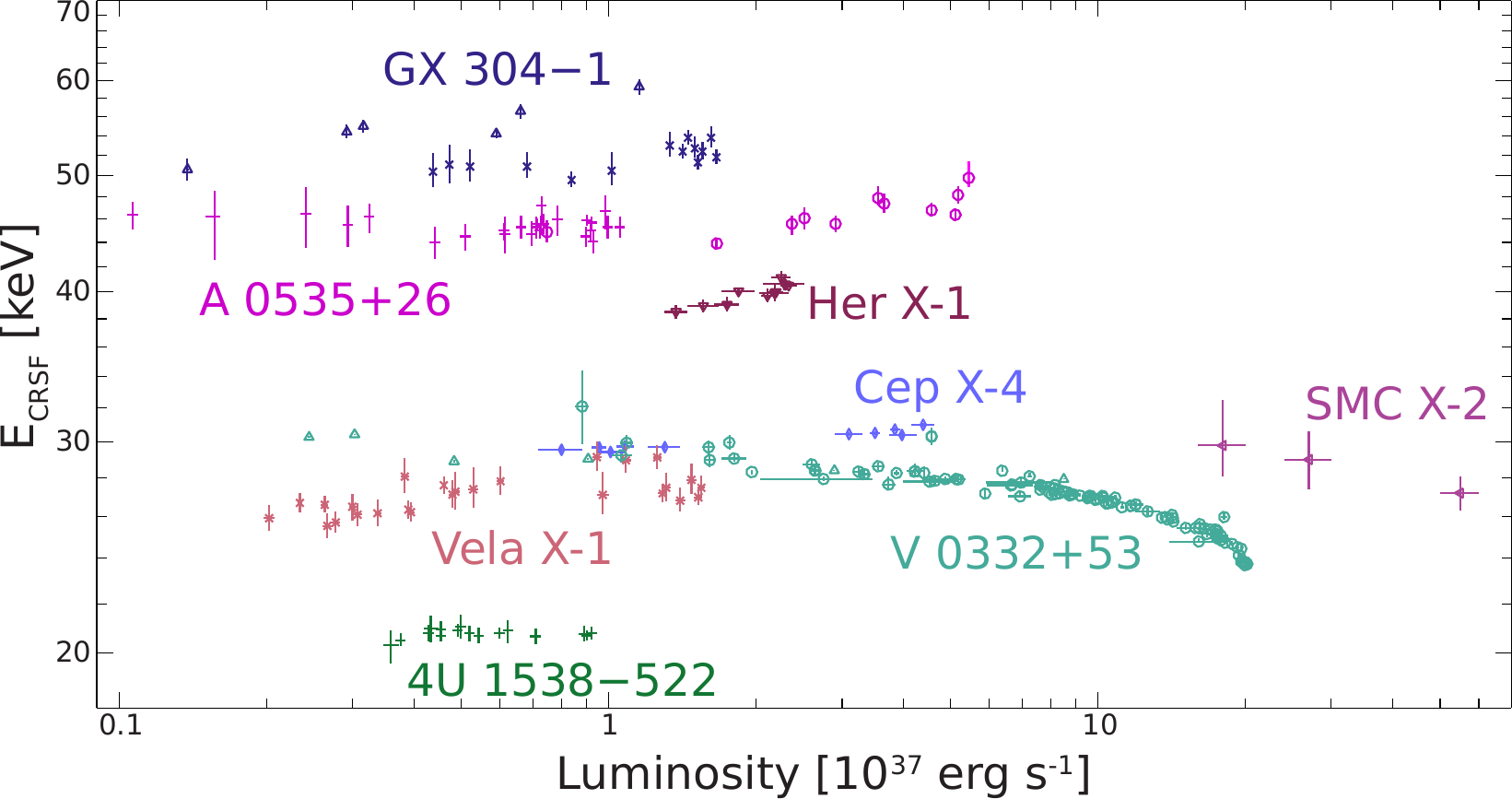}
\hfill
\includegraphics[height=9\baselineskip]{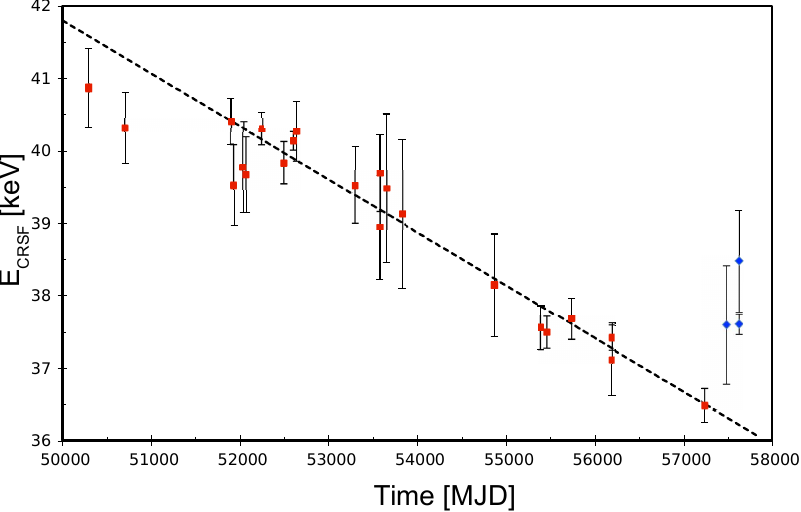}
\vspace{-2ex}
\caption{Left: CRSF energy vs. luminosity for a number of sources, 
adapted from \citet{staubert_cyclotron_2019}. 
Right: Her~X-1's long-term decay in CRSF energy, adapted 
from \citet{staubert_inversion_2017}.}
\label{fig:crsfs}
\vspace{-2ex} 
\end{figure}
The first source with a clear correlation between the CRSF 
energy and the X-ray luminosity was Her~X-1 \citep{staubert_discovery_2007}. 
Since then, measurements of the energy-luminosity relationship have 
been made for several other X-ray pulsars, thanks in large part 
to the the capabilities of \nustar. 
As shown in the left panel of Fig.~\ref{fig:crsfs}, the CRSF energy-luminosity 
relation is bimodal with source luminosity \citep{becker_spectral_2012}.
The two sources with negative CRSF energy-luminosity relations, 
V~0332+53 \citep{tsygankov_v0332_2006} and SMC~X-2 \citep{jaisawal_detection_2017}, 
are the brightest known CRSF sources, while Her~X-1 and the 
other sources displaying positive correlations \citep[e.g., Vela~X-1][]{furst_nustar_2014} 
are lower-luminosity. 
Others at still lower luminosity \citep[e.g., 4U~1538$-$522,][]{hemphill_evidence_2016} 
show no evolution of CRSF energy with luminosity.
Different theoretical models have been put forward to explain these 
dependencies --- e.g. the shock-height model of \citet{becker_spectral_2012}, 
or the reflection-model of \citet{postnov_dependence_2015}.
However, a full explanation is lacking, and is hampered by the small 
number of sources with good enough data to observe a trend, 
particularly at low luminosities. Future observations made with 
detectors with enhanced spectral and temporal resolution, combined 
with new theoretical work, are necessary to understand this 
aspect of accretion onto the magnetic poles of NSs.

\subsection{Evolution of the Continuum Shape with Luminosity}

The observed spectra of many accreting NSs seem to harden with 
increasing luminosity \citep[see, e.g.,][]{reig_nespoli_2013}. 
This behavior can be explained by an increase in 
the ``bulk Comptonization'' process \citep{postnov_dependence_2015}, 
which transfers bulk kinetic energy from the inflowing gas 
directly to the emergent hard X-ray spectra. 
The typical luminosities associated with this type of spectral 
hardening are on the order of $10^{36}-10^{37}$\,erg\,s$^{-1}$. 
There is growing evidence from detailed spectral analyses of 
large numbers of observations that at higher luminosities the 
hardening saturates or even starts to decrease 
again \citep[see, e.g.,][]{postnov_dependence_2015,reig_1901_2016,kuehnel_gro1008_2017}. 
This has been interpreted as a transition from the 
Coulomb-dominated accretion regime to the radiation-dominated one. 
At luminosities well below $10^{36}$\,erg\,s$^{-1}$ very soft 
spectra have been observed, which might indicate a transition 
to a regime where the physical mechanism capable of stopping 
the infalling material is yet to be 
explored \citep{kuehnel_gro1008_2017,tsygankov_gx304_2019}.

\subsection{A New Development: Long-Term Variation of the Cyclotron Line}  \label{sec:crsf_time}
A major discovery of the past decade is the measurement 
by \citet{staubert_long-term_2014} of an upward jump followed by 
a significant secular decay (${\sim}0.3$\,keV\,yr$^{-1}$) in 
the CRSF energy of Her~X-1, unrelated to the source luminosity 
(see Fig.~\ref{fig:crsfs}, right). 
Since then, similar trends have been detected in three more sources: Vela~X-1 \citep[long-term decay, see][]{Laparola_BAT_2016,ji_long-term_2019}, V~0332+53 \citep[short- and long-term changes, see][]{vybornov_changes_2018}, 
and 4U~1538$-$522 \citep[upwards jump, see][]{hemphill_evidence_2016,Hemphill_nustar_2019}. 
The physical mechanism behind these changes is unknown, but must involve some 
restructuring of the accretion flow near the polar regions of the NS ---
e.g., accumulations of accreted plasma could distort the field, or 
the accreted material could collapse or leak out of the 
magnetically-confined mound on the surface. 
This reinforces the need for detailed theoretical and 
simulation-based treatments of NS accretion.

\subsection{Going to Extremes: the Brightest Neutron Stars in the Sky} \label{sec:ulx}

With the recent discovery of ultra-luminous X-ray pulsars (ULXPs), it 
has been shown that NSs can reach isotropic luminosities up to at 
least $10^{41}\,\mathrm{erg}\,\mathrm{s}^{-1}$ 
\citep{bachetti_ultraluminous_2014, israel_accreting_2017}.
Most ULXPs are extragalactic sources, and their large distances make 
detailed studies with existing instruments difficult. 
However, even within the Milky Way and the Magellanic Clouds, we 
have a few examples of NSs breaching their Eddington limit, 
albeit by a less extreme factor. 
The transient Swift~J0243.6+6124 recently underwent an extremely 
bright outburst, likely reaching over $10^{39}\,\mathrm{erg}\,\mathrm{s}^{-1}$ 
at its peak \citep{wilsonhodge_NICER_2018}. 
The source's pulse profile evolved dramatically over the outburst, suggesting 
a changing emission geometry due to transitions between accretion regimes. 
Detailed studies of such transitions are key to understanding the physics involved. 
Another example is LMC~X-4, in which, as discovered by \citet{brumback_discovery_2018}, 
pulsations sometimes disappear without obvious changes in flux, similar 
to the ULXP M82~X-2 \citep{karino_16}. 
The reasons for these changes are still unclear.
To advance our understanding of this behavior, dedicated monitoring 
by instruments with high time resolution, and good spectral and angular 
resolution, is necessary. 
Additionally, ULXPs' apparent super-Eddington accretion rates reinforce the 
need, as laid out in Section~\ref{sec:physics}, for a solid theoretical 
understanding of the emission geometry and the spectral formation 
mechanism operative in such systems.

\section{Requirements on Future X-ray Missions} \label{sec:require}

The topics laid out above allow us to form a fairly clear picture of how X-ray 
missions need to evolve in order to achieve our goals.
Of prime importance is broad-band coverage with good energy resolution, as will 
be provided by missions like \textit{HEX-P} \citep{madsen_hexp_2017}, which 
is necessary to detect CRSFs and constrain their profiles. 
All-sky or wide-field monitoring, flexible scheduling, and fast response times, 
with a mission such as \strobex \citep{ray_strobex_2019}, will be essential 
for responding to sources in outburst. 
We must also be able to track sources to very high and very low luminosities 
in order to probe the full range of accretion regimes. 
Finally, CRSFs are a prime target for X-ray 
polarimetry studies, since large variations in polarization fraction and 
angle are expected over the pulse \citep[see, e.g,][for the case of the Crab pulsar]{vadawale_polarimetry_2018}, 
especially in the vicinity of the CRSF energy. 
With the coming launch of \textit{IXPE}, X-ray polarimetry at softer 
X-ray energies will become a reality, but there is an urgent need for 
an instrument with polarimetry capabilities in the 10--200\,keV band. 
We summarize these mission requirements in Table~\ref{tab:requirements}.

\begin{table}
\caption{Requirements for future X-ray missions\label{tab:requirements}}
\footnotesize
\begin{tabu} to \textwidth {X[0.8,l] X[l] X[l] }
\toprule
Parameter & Requirement & Scientific Justification \\
\midrule
Energy bandpass & 10--200\,keV  & Known CRSF band \\
                & 0.5--10\,keV  & Continuum \& absorption constraint \\
Energy resolution & $\sim 600$\,eV \, @ 60 keV & Detailed CRSF profiles\\
Time resolution & $\,{\lesssim}100$\,$\mu$s & \mbox{Phase-connected, -resolved spectroscopy} \\
Sensitivity & $10^{-13}$\,erg\,cm$^{-2}$\,s$^{-1}$, 6-60 keV, in 1\,ks & Low-$\dot{M}$ and quiescent sources \\
Photon pile-up & ${\lesssim}10$\% @ 5\,Crab & Bright sources in outburst \\
Observing flexibility & ${\sim}1$-day repoint & Rapid response to X-ray transients \\
Polarization & MDP $\lesssim$ 20\% @ 30\,keV & Detect polarization in CRSF band\\
All-sky monitoring & $10^{-10}$\,erg\,cm$^{-2}$\,s$^{-1}$ & Long-timescale variability \\
\bottomrule
\end{tabu}
\end{table}

\pagebreak

\section{References}
\renewcommand{\bibsection}{}
\renewcommand{\bibsep}{0pt}
\begin{multicols}{2}
\bibliography{refs}

\begin{thebibliography}{47}
\providecommand{\natexlab}[1]{#1}
\providecommand{\url}[1]{\texttt{#1}}
\expandafter\ifx\csname urlstyle\endcsname\relax
  \providecommand{\doi}[1]{doi: #1}\else
  \providecommand{\doi}{doi: \begingroup \urlstyle{rm}\Url}\fi

\bibitem[{Abbott} et~al.(2017){Abbott}, {Abbott}, {Abbott}, {Acernese},
  {Ackley}, {Adams}, {Adams}, {Addesso}, {Adhikari}, {Adya}, and
  et~al.]{abbott_multimessenger_2017}
B.~P. {Abbott}, R.~{Abbott}, T.~D. {Abbott}, F.~{Acernese}, K.~{Ackley},
  C.~{Adams}, T.~{Adams}, P.~{Addesso}, R.~X. {Adhikari}, V.~B. {Adya}, and
  et~al.
\newblock {Multi-messenger Observations of a Binary Neutron Star Merger}.
\newblock \emph{\apjl}, 848:\penalty0 L12, Oct. 2017.
\newblock \doi{10.3847/2041-8213/aa91c9}.

\bibitem[Araya and Harding(1996)]{araya_cylotron_1996}
R.~A. Araya and A.~K. Harding.
\newblock Cylotron line models for the {{X}}-ray pulsar {{A}} 0535+26.
\newblock \emph{Astronomy and Astrophysics Supplement Series}, 120, Nov. 1996.

\bibitem[Arons et~al.(1987)Arons, Klein, and Lea]{arons_radiation_1987}
J.~Arons, R.~I. Klein, and S.~M. Lea.
\newblock Radiation gasdynamics of polar {{CAP}} accretion onto magnetized
  neutron stars - {{Basic}} theory.
\newblock \emph{The Astrophysical Journal}, 312:\penalty0 666--699, Jan. 1987.
\newblock ISSN 0004-637X.
\newblock \doi{10.1086/164912}.

\bibitem[{Bachetti} et~al.(2014){Bachetti}, {Harrison}, {Walton},
  {Grefenstette}, {Chakrabarty}, {F{\"u}rst}, {Barret}, {Beloborodov}, {Boggs},
  {Christensen}, {Craig}, {Fabian}, {Hailey}, {Hornschemeier}, {Kaspi},
  {Kulkarni}, {Maccarone}, {Miller}, {Rana}, {Stern}, {Tendulkar}, {Tomsick},
  {Webb}, and {Zhang}]{bachetti_ultraluminous_2014}
M.~{Bachetti}, F.~A. {Harrison}, D.~J. {Walton}, B.~W. {Grefenstette},
  D.~{Chakrabarty}, F.~{F{\"u}rst}, D.~{Barret}, A.~{Beloborodov}, S.~E.
  {Boggs}, F.~E. {Christensen}, W.~W. {Craig}, A.~C. {Fabian}, C.~J. {Hailey},
  A.~{Hornschemeier}, V.~{Kaspi}, S.~R. {Kulkarni}, T.~{Maccarone}, J.~M.
  {Miller}, V.~{Rana}, D.~{Stern}, S.~P. {Tendulkar}, J.~{Tomsick}, N.~A.
  {Webb}, and W.~W. {Zhang}.
\newblock {An ultraluminous X-ray source powered by an accreting neutron star}.
\newblock \emph{\nat}, 514:\penalty0 202--204, Oct. 2014.
\newblock \doi{10.1038/nature13791}.

\bibitem[Basko and Sunyaev(1976)]{basko_limiting_1976}
M.~M. Basko and R.~A. Sunyaev.
\newblock The {{Limiting Luminosity}} of {{Accreting Neutron Stars With
  Magnetic Fields}}.
\newblock \emph{Monthly Notices of the Royal Astronomical Society},
  175\penalty0 (2):\penalty0 395--417, Jan. 1976.
\newblock ISSN 0035-8711, 1365-2966.
\newblock \doi{10.1093/mnras/175.2.395}.

\bibitem[Becker and Kazanas(2001)]{becker_exact_2001}
P.~A. Becker and D.~Kazanas.
\newblock Exact {{Expressions}} for the {{Critical Mach Numbers}} in the
  {{Two}}-{{Fluid Model}} of {{Cosmic}}-{{Ray}}-modified {{Shocks}}.
\newblock \emph{The Astrophysical Journal}, 546:\penalty0 429--446, Jan. 2001.
\newblock ISSN 0004-637X.
\newblock \doi{10.1086/318257}.

\bibitem[Becker and Wolff(2007)]{becker_thermal_2007}
P.~A. Becker and M.~T. Wolff.
\newblock Thermal and {{Bulk Comptonization}} in {{Accretion}}-powered
  {{X}}-{{Ray Pulsars}}.
\newblock \emph{The Astrophysical Journal}, 654\penalty0 (1):\penalty0 435,
  2007.
\newblock \doi{10.1086/509108}.

\bibitem[Becker et~al.(2012)Becker, Klochkov, Sch\"onherr, Nishimura, Ferrigno,
  Caballero, Kretschmar, Wolff, Wilms, and Staubert]{becker_spectral_2012}
P.~A. Becker, D.~Klochkov, G.~Sch\"onherr, O.~Nishimura, C.~Ferrigno,
  I.~Caballero, P.~Kretschmar, M.~T. Wolff, J.~Wilms, and R.~Staubert.
\newblock Spectral formation in accreting {{X}}-ray pulsars: Bimodal variation
  of the cyclotron energy with luminosity.
\newblock \emph{Astronomy and Astrophysics}, 544:\penalty0 A123, Aug. 2012.
\newblock ISSN 0004-6361.
\newblock \doi{10.1051/0004-6361/201219065}.

\bibitem[{Brumback} et~al.(2018){Brumback}, {Hickox}, {Bachetti}, {Ballhausen},
  {F{\"u}rst}, {Pike}, {Pottschmidt}, {Tomsick}, and
  {Wilms}]{brumback_discovery_2018}
M.~C. {Brumback}, R.~C. {Hickox}, M.~{Bachetti}, R.~{Ballhausen}, F.~S.
  {F{\"u}rst}, S.~{Pike}, K.~{Pottschmidt}, J.~A. {Tomsick}, and J.~{Wilms}.
\newblock {Discovery of Pulsation Dropout and Turn-on during the High State of
  the Accreting X-Ray Pulsar LMC X-4}.
\newblock \emph{\apjl}, 861:\penalty0 L7, July 2018.
\newblock \doi{10.3847/2041-8213/aacd13}.

\bibitem[Davidson(1973)]{davidson_accretion_1973}
K.~Davidson.
\newblock Accretion at a {{Magnetic Pole}} of a {{Neutron Star}}.
\newblock \emph{Nature Physical Science}, 246\penalty0 (149):\penalty0 1--4,
  Nov. 1973.
\newblock ISSN 0300-8746.
\newblock \doi{10.1038/physci246001a0}.

\bibitem[Farinelli et~al.(2012)Farinelli, Ceccobello, Romano, and
  Titarchuk]{farinelli_numerical_2012}
R.~Farinelli, C.~Ceccobello, P.~Romano, and L.~Titarchuk.
\newblock Numerical solution of the radiative transfer equation: {{X}}-ray
  spectral formation from cylindrical accretion onto a magnetized neutron star.
\newblock \emph{Astronomy and Astrophysics}, 538:\penalty0 A67, Feb. 2012.
\newblock ISSN 0004-6361.
\newblock \doi{10.1051/0004-6361/201118008}.

\bibitem[F\"urst et~al.(2014)F\"urst, Pottschmidt, Wilms, Tomsick, Bachetti,
  Boggs, Christensen, Craig, Grefenstette, Hailey, Harrison, Madsen, Miller,
  Stern, Walton, and Zhang]{furst_nustar_2014}
F.~F\"urst, K.~Pottschmidt, J.~Wilms, J.~A. Tomsick, M.~Bachetti, S.~E. Boggs,
  F.~E. Christensen, W.~W. Craig, B.~W. Grefenstette, C.~J. Hailey,
  F.~Harrison, K.~K. Madsen, J.~M. Miller, D.~Stern, D.~J. Walton, and
  W.~Zhang.
\newblock {{NuSTAR Discovery}} of a {{Luminosity Dependent Cyclotron Line
  Energy}} in {{Vela X}}-1.
\newblock \emph{The Astrophysical Journal}, 780\penalty0 (2):\penalty0 133,
  2014.
\newblock ISSN 0004-637X.
\newblock \doi{10.1088/0004-637X/780/2/133}.

\bibitem[Ghosh and Lamb(1978)]{ghosh_disk_1978}
P.~Ghosh and F.~K. Lamb.
\newblock Disk accretion by magnetic neutron stars.
\newblock \emph{The Astrophysical Journal}, 223:\penalty0 L83, July 1978.
\newblock \doi{10.1086/182734}.

\bibitem[Hemphill et~al.(2016)Hemphill, Rothschild, F\"urst, Grinberg,
  Klochkov, Kretschmar, Pottschmidt, Staubert, and
  Wilms]{hemphill_evidence_2016}
P.~B. Hemphill, R.~E. Rothschild, F.~F\"urst, V.~Grinberg, D.~Klochkov,
  P.~Kretschmar, K.~Pottschmidt, R.~Staubert, and J.~Wilms.
\newblock Evidence for an evolving cyclotron line energy in {{4U}} 1538-522.
\newblock \emph{Monthly Notices of the Royal Astronomical Society},
  458\penalty0 (3):\penalty0 2745, 2016.
\newblock \doi{10.1093/mnras/stw470}.

\bibitem[{Hemphill} et~al.(2019){Hemphill}, {Rothschild}, {Cheatham},
  {F{\"u}rst}, {Kretschmar}, {K{\"u}hnel}, {Pottschmidt}, {Staubert}, {Wilms},
  and {Wolff}]{Hemphill_nustar_2019}
P.~B. {Hemphill}, R.~E. {Rothschild}, D.~M. {Cheatham}, F.~{F{\"u}rst},
  P.~{Kretschmar}, M.~{K{\"u}hnel}, K.~{Pottschmidt}, R.~{Staubert},
  J.~{Wilms}, and M.~T. {Wolff}.
\newblock {The First NuSTAR Observation of 4U 1538-522: Updated Orbital
  Ephemeris and A Strengthened Case for an Evolving Cyclotron Line Energy}.
\newblock \emph{arXiv e-prints}, art. arXiv:1901.11071, Jan 2019.

\bibitem[{Israel} et~al.(2017){Israel}, {Belfiore}, {Stella}, {Esposito},
  {Casella}, {De Luca}, {Marelli}, {Papitto}, {Perri}, {Puccetti}, {Castillo},
  {Salvetti}, {Tiengo}, {Zampieri}, {D'Agostino}, {Greiner}, {Haberl},
  {Novara}, {Salvaterra}, {Turolla}, {Watson}, {Wilms}, and
  {Wolter}]{israel_accreting_2017}
G.~L. {Israel}, A.~{Belfiore}, L.~{Stella}, P.~{Esposito}, P.~{Casella}, A.~{De
  Luca}, M.~{Marelli}, A.~{Papitto}, M.~{Perri}, S.~{Puccetti}, G.~A.~R.
  {Castillo}, D.~{Salvetti}, A.~{Tiengo}, L.~{Zampieri}, D.~{D'Agostino},
  J.~{Greiner}, F.~{Haberl}, G.~{Novara}, R.~{Salvaterra}, R.~{Turolla},
  M.~{Watson}, J.~{Wilms}, and A.~{Wolter}.
\newblock {An accreting pulsar with extreme properties drives an ultraluminous
  x-ray source in NGC 5907}.
\newblock \emph{Science}, 355:\penalty0 817--819, Feb. 2017.
\newblock \doi{10.1126/science.aai8635}.

\bibitem[{Jaisawal} and {Naik}(2016)]{jaisawal_detection_2017}
G.~K. {Jaisawal} and S.~{Naik}.
\newblock {Detection of cyclotron resonance scattering feature in high-mass
  X-ray binary pulsar SMC X-2}.
\newblock \emph{\mnras}, 461:\penalty0 L97--L101, Sept. 2016.
\newblock \doi{10.1093/mnrasl/slw108}.

\bibitem[{Ji} et~al.(2019){Ji}, {Staubert}, {Ducci}, {Santangelo}, {Zhang}, and
  {Chang}]{ji_long-term_2019}
L.~{Ji}, R.~{Staubert}, L.~{Ducci}, A.~{Santangelo}, S.~{Zhang}, and
  Z.~{Chang}.
\newblock {Long-term evolutions of the cyclotron line energies in Her X-1, Vela
  X-1, and Cen X-3 as observed with Swift/BAT}.
\newblock \emph{\mnras}, 484:\penalty0 3797--3805, Apr. 2019.
\newblock \doi{10.1093/mnras/stz264}.

\bibitem[{Karino} and {Miller}(2016)]{karino_16}
S.~{Karino} and J.~C. {Miller}.
\newblock {Accretion mode of the ultraluminous X-ray source M82 X-2}.
\newblock \emph{\mnras}, 462:\penalty0 3476--3482, Nov. 2016.
\newblock \doi{10.1093/mnras/stw1180}.

\bibitem[{K{\"u}hnel} et~al.(2017){K{\"u}hnel}, {F{\"u}rst}, {Pottschmidt},
  {Kreykenbohm}, {Ballhausen}, {Falkner}, {Rothschild}, {Klochkov}, and
  {Wilms}]{kuehnel_gro1008_2017}
M.~{K{\"u}hnel}, F.~{F{\"u}rst}, K.~{Pottschmidt}, I.~{Kreykenbohm},
  R.~{Ballhausen}, S.~{Falkner}, R.~E. {Rothschild}, D.~{Klochkov}, and
  J.~{Wilms}.
\newblock Evidence for different accretion regimes in gro j1008-57.
\newblock \emph{Astronomy \& Astrophysics}, 607:\penalty0 A88, Nov. 2017.

\bibitem[{La Parola} et~al.(2016){La Parola}, {Cusumano}, {Segreto}, and
  {D'A{\`\i}}]{Laparola_BAT_2016}
V.~{La Parola}, G.~{Cusumano}, A.~{Segreto}, and A.~{D'A{\`\i}}.
\newblock {The Swift-BAT monitoring reveals a long-term decay of the cyclotron
  line energy in Vela X-1}.
\newblock \emph{\mnras}, 463:\penalty0 185--190, Nov 2016.
\newblock \doi{10.1093/mnras/stw1915}.

\bibitem[Langer and Rappaport(1982)]{langer_low-luminosity_1982}
S.~H. Langer and S.~Rappaport.
\newblock Low-luminosity accretion onto magnetized neutron stars.
\newblock \emph{The Astrophysical Journal}, 257:\penalty0 733, June 1982.
\newblock ISSN 0004-637X, 1538-4357.
\newblock \doi{10.1086/160028}.

\bibitem[{Lattimer}(2012)]{lattimer_2012}
J.~M. {Lattimer}.
\newblock The nuclear equation of state and neutron star masses.
\newblock \emph{Annual Review of Nuclear and Particle Science}, 62:\penalty0
  485--515, Nov. 2012.

\bibitem[{Madsen} et~al.(2017){Madsen}, {Harrison}, {Stern}, {Grefenstette},
  {Rana}, and {Miyasaka}]{madsen_hexp_2017}
K.~{Madsen}, F.~{Harrison}, D.~{Stern}, B.~{Grefenstette}, V.~{Rana}, and
  H.~{Miyasaka}.
\newblock {The High-Energy X-ray Probe (HEX-P)}.
\newblock In \emph{AAS/High Energy Astrophysics Division \#16}, volume~16 of
  \emph{AAS/High Energy Astrophysics Division}, page 103.21, Aug. 2017.

\bibitem[Meszaros(1992)]{meszaros_high-energy_1992}
P.~Meszaros.
\newblock \emph{High-energy radiation from magnetized neutron stars}.
\newblock University of Chicago Press, 1992.

\bibitem[Postnov et~al.(2015)Postnov, Gornostaev, Klochkov, Laplace, Lukin, and
  Shakura]{postnov_dependence_2015}
K.~A. Postnov, M.~I. Gornostaev, D.~Klochkov, E.~Laplace, V.~V. Lukin, and
  N.~I. Shakura.
\newblock On the dependence of the {{X}}-ray continuum variations with
  luminosity in accreting {{X}}-ray pulsars.
\newblock \emph{Monthly Notices of the Royal Astronomical Society},
  452:\penalty0 1601--1611, Sept. 2015.
\newblock ISSN 0035-8711.
\newblock \doi{10.1093/mnras/stv1393}.

\bibitem[{Ray} et~al.(2019){Ray}, {Arzoumanian}, {Ballantyne}, {Bozzo},
  {Brandt}, {Brenneman}, {Chakrabarty}, {Christophersen}, {DeRosa}, {Feroci},
  {Gendreau}, {Goldstein}, {Hartmann}, {Hernanz}, {Jenke}, {Kara}, {Maccarone},
  {McDonald}, {Nowak}, {Phlips}, {Remillard}, {Stevens}, {Tomsick}, {Watts},
  {Wilson-Hodge}, {Wood}, and {Zane}]{ray_strobex_2019}
P.~S. {Ray}, Z.~{Arzoumanian}, D.~{Ballantyne}, E.~{Bozzo}, S.~{Brandt},
  L.~{Brenneman}, D.~{Chakrabarty}, M.~{Christophersen}, A.~r. {DeRosa},
  M.~{Feroci}, K.~{Gendreau}, A.~{Goldstein}, D.~{Hartmann}, M.~{Hernanz},
  P.~{Jenke}, E.~{Kara}, T.~{Maccarone}, M.~{McDonald}, M.~{Nowak},
  B.~{Phlips}, R.~{Remillard}, A.~{Stevens}, J.~{Tomsick}, A.~{Watts},
  C.~{Wilson-Hodge}, K.~{Wood}, and S.~{Zane}.
\newblock {STROBE-X: X-ray Timing and Spectroscopy on Dynamical Timescales from
  Microseconds to Years}.
\newblock \emph{arXiv e-prints}, art. arXiv:1903.03035, Mar 2019.

\bibitem[{Reig} and {Milonaki}(2016)]{reig_1901_2016}
P.~{Reig} and F.~{Milonaki}.
\newblock {Accretion regimes in the X-ray pulsar 4U 1901+03}.
\newblock \emph{Astronomy \& Astrophysics}, 594:\penalty0 A45, Oct. 2016.

\bibitem[{Reig} and {Nespoli}(2013)]{reig_nespoli_2013}
P.~{Reig} and E.~{Nespoli}.
\newblock Patterns of variability in be/x-ray pulsars during giant outbursts.
\newblock \emph{Astronomy \& Astrophysics}, 551, Mar. 2013.

\bibitem[Sch\"onherr et~al.(2007)Sch\"onherr, Wilms, Kretschmar, Kreykenbohm,
  Santangelo, Rothschild, Coburn, and Staubert]{schonherr_model_2007}
G.~Sch\"onherr, J.~Wilms, P.~Kretschmar, I.~Kreykenbohm, A.~Santangelo, R.~E.
  Rothschild, W.~Coburn, and R.~Staubert.
\newblock A model for cyclotron resonance scattering features.
\newblock \emph{Astronomy and Astrophysics}, 472:\penalty0 353--365, Sept.
  2007.
\newblock ISSN 0004-6361.
\newblock \doi{10.1051/0004-6361:20077218}.

\bibitem[Sch\"onherr et~al.(2014)Sch\"onherr, Schwarm, Falkner, Dauser,
  Ferrigno, K\"uhnel, Klochkov, Kretschmar, Becker, Wolff, Pottschmidt,
  Falanga, Kreykenbohm, F\"urst, Staubert, and Wilms]{schonherr_formation_2014}
G.~Sch\"onherr, F.-W. Schwarm, S.~Falkner, T.~Dauser, C.~Ferrigno, M.~K\"uhnel,
  D.~Klochkov, P.~Kretschmar, P.~A. Becker, M.~T. Wolff, K.~Pottschmidt,
  M.~Falanga, I.~Kreykenbohm, F.~F\"urst, R.~Staubert, and J.~Wilms.
\newblock Formation of phase lags at the cyclotron energies in the pulse
  profiles of magnetized, accreting neutron stars.
\newblock \emph{Astronomy \& Astrophysics}, 564:\penalty0 L8, Apr. 2014.
\newblock ISSN 0004-6361, 1432-0746.
\newblock \doi{10.1051/0004-6361/201322448}.

\bibitem[Schwarm et~al.(2017{\natexlab{a}})Schwarm, Ballhausen, Falkner,
  Sch\"onherr, Pottschmidt, Wolff, Becker, F\"urst, {Marcu-Cheatham}, Hemphill,
  {Sokolova-Lapa}, Dauser, Klochkov, Ferrigno, and
  Wilms]{schwarm_cyclotron_2017b}
F.-W. Schwarm, R.~Ballhausen, S.~Falkner, G.~Sch\"onherr, K.~Pottschmidt, M.~T.
  Wolff, P.~A. Becker, F.~F\"urst, D.~M. {Marcu-Cheatham}, P.~B. Hemphill,
  E.~{Sokolova-Lapa}, T.~Dauser, D.~Klochkov, C.~Ferrigno, and J.~Wilms.
\newblock Cyclotron resonant scattering feature simulations. {{II}}.
  {{Description}} of the {{CRSF}} simulation process.
\newblock \emph{Astronomy and Astrophysics}, 601:\penalty0 A99, May
  2017{\natexlab{a}}.
\newblock \doi{10.1051/0004-6361/201630250}.

\bibitem[Schwarm et~al.(2017{\natexlab{b}})Schwarm, Sch\"onherr, Falkner,
  Pottschmidt, Wolff, Becker, {Sokolova-Lapa}, Klochkov, Ferrigno, F\"urst,
  Hemphill, {Marcu-Cheatham}, Dauser, and Wilms]{schwarm_cyclotron_2017a}
F.-W. Schwarm, G.~Sch\"onherr, S.~Falkner, K.~Pottschmidt, M.~T. Wolff, P.~A.
  Becker, E.~{Sokolova-Lapa}, D.~Klochkov, C.~Ferrigno, F.~F\"urst, P.~B.
  Hemphill, D.~M. {Marcu-Cheatham}, T.~Dauser, and J.~Wilms.
\newblock Cyclotron resonant scattering feature simulations. {{I}}.
  {{Thermally}} averaged cyclotron scattering cross sections, mean free
  photon-path tables, and electron momentum sampling.
\newblock \emph{Astronomy and Astrophysics}, 597:\penalty0 A3, Jan.
  2017{\natexlab{b}}.
\newblock \doi{10.1051/0004-6361/201629352}.

\bibitem[Shakura and Sunyaev(1973)]{shakura_black_1973}
N.~I. Shakura and R.~A. Sunyaev.
\newblock Black holes in binary systems. {{Observational}} appearance.
\newblock \emph{Astronomy and Astrophysics}, 24, 1973.

\bibitem[Staubert et~al.(2007)Staubert, Shakura, Postnov, Wilms, Rothschild,
  Coburn, Rodina, and Klochkov]{staubert_discovery_2007}
R.~Staubert, N.~I. Shakura, K.~Postnov, J.~Wilms, R.~E. Rothschild, W.~Coburn,
  L.~Rodina, and D.~Klochkov.
\newblock Discovery of a flux-related change of the cyclotron line energy in
  {{Hercules X}}-1.
\newblock \emph{Astronomy and Astrophysics}, 465:\penalty0 L25--L28, Apr. 2007.
\newblock ISSN 0004-6361.
\newblock \doi{10.1051/0004-6361:20077098}.

\bibitem[Staubert et~al.(2014)Staubert, Klochkov, Wilms, Postnov, Shakura,
  Rothschild, F\"urst, and Harrison]{staubert_long-term_2014}
R.~Staubert, D.~Klochkov, J.~Wilms, K.~Postnov, N.~I. Shakura, R.~E.
  Rothschild, F.~F\"urst, and F.~A. Harrison.
\newblock Long-term change in the cyclotron line energy in {{Hercules X}}-1.
\newblock \emph{Astronomy and Astrophysics}, 572:\penalty0 A119, Dec. 2014.
\newblock ISSN 0004-6361.
\newblock \doi{10.1051/0004-6361/201424203}.

\bibitem[Staubert et~al.(2017)Staubert, Klochkov, F\"urst, Wilms, Rothschild,
  and Harrison]{staubert_inversion_2017}
R.~Staubert, D.~Klochkov, F.~F\"urst, J.~Wilms, R.~E. Rothschild, and
  F.~Harrison.
\newblock Inversion of the decay of the cyclotron line energy in {{Hercules
  X}}-1.
\newblock \emph{Astronomy and Astrophysics}, 606:\penalty0 L13, Oct. 2017.
\newblock \doi{10.1051/0004-6361/201731927}.

\bibitem[Staubert et~al.(2019)Staubert, Tr\"umper, Kendziorra, Klochkov,
  Postnov, Kretschmar, Pottschmidt, Haberl, Rothschild, Santangelo, Wilms,
  Kreykenbohm, and F\"urst]{staubert_cyclotron_2019}
R.~Staubert, J.~Tr\"umper, E.~Kendziorra, D.~Klochkov, K.~Postnov,
  P.~Kretschmar, K.~Pottschmidt, F.~Haberl, R.~E. Rothschild, A.~Santangelo,
  J.~Wilms, I.~Kreykenbohm, and F.~F\"urst.
\newblock Cyclotron lines in highly magnetized neutron stars.
\newblock \emph{Astronomy \& Astrophysics}, 622:\penalty0 A61, Jan. 2019.
\newblock \doi{10.1051/0004-6361/201834479}.

\bibitem[Tsygankov et~al.(2006)Tsygankov, Lutovinov, Churazov, and
  Sunyaev]{tsygankov_v0332_2006}
S.~S. Tsygankov, A.~A. Lutovinov, E.~M. Churazov, and R.~A. Sunyaev.
\newblock V0332+53 in the outburst of 2004-2005: Luminosity dependence of the
  cyclotron line and pulse profile.
\newblock \emph{Monthly Notices of the Royal Astronomical Society},
  371\penalty0 (1):\penalty0 19, Sept. 2006.
\newblock \doi{10.1111/j.1365-2966.2006.10610.x}.

\bibitem[{Tsygankov} et~al.(2019){Tsygankov}, {Rouco Escorial}, {Suleimanov},
  {Mushtukov}, {Doroshenko}, {Lutovinov}, {Wijnands}, and
  {Poutanen}]{tsygankov_gx304_2019}
S.~S. {Tsygankov}, A.~{Rouco Escorial}, V.~F. {Suleimanov}, A.~A. {Mushtukov},
  V.~{Doroshenko}, A.~A. {Lutovinov}, R.~{Wijnands}, and J.~{Poutanen}.
\newblock Dramatic spectral transition of x-ray pulsar gx 304-1 in low luminous
  state.
\newblock \emph{Monthly Notices of the Royal Astronomical Society},
  483:\penalty0 L144--L148, Feb. 2019.

\bibitem[{Vadawale} et~al.(2018){Vadawale}, {Chattopadhyay}, {Mithun}, {Rao},
  {Bhattacharya}, {Vibhute}, {Bhalerao}, {Dewangan}, {Misra}, {Paul}, {Basu},
  {Joshi}, {Sreekumar}, {Samuel}, {Priya}, {Vinod}, and
  {Seetha}]{vadawale_polarimetry_2018}
S.~V. {Vadawale}, T.~{Chattopadhyay}, N.~P.~S. {Mithun}, A.~R. {Rao},
  D.~{Bhattacharya}, A.~{Vibhute}, V.~B. {Bhalerao}, G.~C. {Dewangan},
  R.~{Misra}, B.~{Paul}, A.~{Basu}, B.~C. {Joshi}, S.~{Sreekumar}, E.~{Samuel},
  P.~{Priya}, P.~{Vinod}, and S.~{Seetha}.
\newblock {Phase-resolved X-ray polarimetry of the Crab pulsar with the
  AstroSat CZT Imager}.
\newblock \emph{Nature Astronomy}, 2:\penalty0 50--55, Nov 2018.
\newblock \doi{10.1038/s41550-017-0293-z}.

\bibitem[Vybornov et~al.(2018)Vybornov, Doroshenko, Staubert, and
  Santangelo]{vybornov_changes_2018}
V.~Vybornov, V.~Doroshenko, R.~Staubert, and A.~Santangelo.
\newblock Changes in the cyclotron line energy on short and long timescales in
  {{V}} 0332+53.
\newblock \emph{Astronomy and Astrophysics}, 610:\penalty0 A88, Mar. 2018.
\newblock ISSN 0004-6361.
\newblock \doi{10.1051/0004-6361/201731750}.

\bibitem[Wang et~al.(1988)Wang, Wasserman, and Salpeter]{wang_cyclotron_1988}
J.~C.~L. Wang, I.~M. Wasserman, and E.~E. Salpeter.
\newblock Cyclotron line resonant transfer through neutron star atmospheres.
\newblock \emph{The Astrophysical Journal Supplement Series}, 68:\penalty0
  735--802, Dec. 1988.
\newblock ISSN 0067-0049.
\newblock \doi{10.1086/191303}.

\bibitem[West et~al.(2017{\natexlab{a}})West, Wolfram, and
  Becker]{west_dynamical_2017}
B.~F. West, K.~D. Wolfram, and P.~A. Becker.
\newblock A {{New Two}}-fluid {{Radiation}}-hydrodynamical {{Model}} for
  {{X}}-{{Ray Pulsar Accretion Columns}}.
\newblock \emph{The Astrophysical Journal}, 835\penalty0 (2):\penalty0 129,
  Feb. 2017{\natexlab{a}}.
\newblock \doi{10.3847/1538-4357/835/2/129}.

\bibitem[West et~al.(2017{\natexlab{b}})West, Wolfram, and
  Becker]{west_spectra_2017}
B.~F. West, K.~D. Wolfram, and P.~A. Becker.
\newblock Dynamical and {{Radiative Properties}} of {{X}}-{{Ray Pulsar
  Accretion Columns}}: {{Phase}}-averaged {{Spectra}}.
\newblock \emph{The Astrophysical Journal}, 835\penalty0 (2):\penalty0 130,
  Feb. 2017{\natexlab{b}}.
\newblock \doi{10.3847/1538-4357/835/2/130}.

\bibitem[{Wilson-Hodge} et~al.(2018){Wilson-Hodge}, {Malacaria}, {Jenke},
  {Jaisawal}, {Kerr}, {Wolff}, {Arzoumanian}, {Chakrabarty}, {Doty},
  {Gendreau}, {Guillot}, {Ho}, {LaMarr}, {Markwardt}, {{\"O}zel}, {Prigozhin},
  {Ray}, {Ramos-Lerate}, {Remillard}, {Strohmayer}, {Vezie}, {Wood}, and {NICER
  Science Team}]{wilsonhodge_NICER_2018}
C.~A. {Wilson-Hodge}, C.~{Malacaria}, P.~A. {Jenke}, G.~K. {Jaisawal},
  M.~{Kerr}, M.~T. {Wolff}, Z.~{Arzoumanian}, D.~{Chakrabarty}, J.~P. {Doty},
  K.~C. {Gendreau}, S.~{Guillot}, W.~C.~G. {Ho}, B.~{LaMarr}, C.~B.
  {Markwardt}, F.~{{\"O}zel}, G.~Y. {Prigozhin}, P.~S. {Ray},
  M.~{Ramos-Lerate}, R.~A. {Remillard}, T.~E. {Strohmayer}, M.~L. {Vezie},
  K.~S. {Wood}, and {NICER Science Team}.
\newblock {NICER and Fermi GBM Observations of the First Galactic Ultraluminous
  X-Ray Pulsar Swift J0243.6+6124}.
\newblock \emph{The Astrophysical Journal}, 863:\penalty0 9, Aug. 2018.
\newblock \doi{10.3847/1538-4357/aace60}.

\bibitem[Wolff et~al.(2016)Wolff, Becker, Gottlieb, F\"urst, Hemphill,
  {Marcu-Cheatham}, Pottschmidt, Schwarm, Wilms, and Wood]{wolff_nustar_2016-1}
M.~T. Wolff, P.~A. Becker, A.~M. Gottlieb, F.~F\"urst, P.~B. Hemphill, D.~M.
  {Marcu-Cheatham}, K.~Pottschmidt, F.-W. Schwarm, J.~Wilms, and K.~S. Wood.
\newblock The {{NuSTAR X}}-{{Ray Spectrum}} of {{Hercules X}}-1: {{A
  Radiation}}-dominated {{Radiative Shock}}.
\newblock \emph{The Astrophysical Journal}, 831\penalty0 (2):\penalty0 194,
  Nov. 2016.
\newblock \doi{10.3847/0004-637X/831/2/194}.

\end{thebibliography}
\end{multicols}

\end{document}